\documentclass[a4paper,11pt, onecolumn]{article}

\usepackage[a4paper]{geometry}		
\usepackage[font=small,labelfont=bf]{caption}

\usepackage{ae} 			
\usepackage[english]{babel} 	
\usepackage{cite}
\usepackage{amsmath} 		
\usepackage{amssymb} 		
\usepackage{bm} 			

\usepackage{float} 			
\usepackage{graphicx} 		
\usepackage{caption}
\usepackage{subcaption}
\usepackage{epstopdf}
\usepackage{sidecap}

\begin{document}

\title{In-medium parton branching beyond eikonal approximation}
%
%

\author{Liliana Apolin\'{a}rio 
	\thanks{\texttt{liliana@lip.pt}} \\ 
	\textit{\small LIP Lisboa, Av. Elias Garcia 14, 1, 1000-149 Lisbon, Portugal}
}
\date{}

\maketitle
%

The description of the in-medium modifications of partonic showers has been at the forefront of current theoretical and experimental efforts in heavy-ion collisions. It provides a unique laboratory to extend our knowledge frontier of the theory of the strong interactions, and to assess the properties of the hot and dense medium (QGP) that is produced in ultra-relativistic heavy-ion collisions at RHIC and the LHC. The theory of jet quenching, a commonly used alias for the modifications of the parton branching resulting from the interactions with the QGP, has been significantly developed over the last years. Within a weak coupling approach, several elementary processes that build up the parton shower evolution, such as single gluon emissions, interference effects between successive emissions and corrections to radiative energy loss of massive quarks, have been addressed both at eikonal accuracy and beyond by taking into account the Brownian motion that high-energy particles experience when traversing a hot and dense medium. In this work, by using the setup of single gluon emission from a color correlated quark-antiquark pair in a singlet state ($q\bar{q}$ antenna), we calculate the in-medium gluon radiation spectrum beyond the eikonal approximation. The results show that we are able to factorize broadening effects from the modifications of the radiation process itself. This constitutes the final proof that a probabilistic picture of the parton shower evolution holds even in the presence of a QGP.

\section{Introduction}
\label{intro}
\par Ultra-relativistic heavy-ion collisions (HIC) are a unique window of opportunity to study, in a controlled environment, a new state of matter called the quark-gluon plasma (QGP). This hot and dense medium can be described using asymptotically free quarks and gluons, the fundamental degrees of freedom of the theory of the strong interactions, the Quantum Chromodynamics (QCD). Therefore, HIC are a perfect laboratory to assess the QCD fundamental properties, such as asymptotic freedom and the Deconfinement/Confinement transition. The observation of this new state of matter has to be made indirectly, through self-generated probes, due to its very short lifetime. Among the several possibilities, hard probes, such as high momentum particles or jets, can provide information on the macroscopic properties of the created medium through the modifications that are observed with respect to extrapolations to proton-proton (pp) collisions - a phenomena that is generically known as \textit{Jet Quenching}. Moreover, the description of leading particle and jet cross-sections has been successfully addressed by perturbative QCD (pQCD) in pp collisions. The new challenge now is to understand at what extent pQCD can be applicable to such dense and hot systems as HIC.
\par Factorization theorem states that:
\begin{equation}
	d\sigma_{(vac)}^{AA \rightarrow h + rest} = \sum_{ijk} f_{i/A} (x_1, Q^2) \otimes f_{j/A} (x_2, Q^2) \otimes \hat{\sigma}_{ij \rightarrow k} \otimes D^{(vac)}_{f \rightarrow h} (z, \mu_F) \, ,
\end{equation}
where $f_{i(j)/A}$ describe the probability of finding a parton $i(j)$ in the incoming hadron with a virtuality $Q$, $\hat{\sigma}_{ij \rightarrow j +k}$ the elementary cross-section and $D_{f \rightarrow h}(z, \mu_F^2)$ the universal hadronisation functions whose definition depends on some factorisation scale, $\mu_F$.
\par Probes originated from a hard process, with a large momentum transfer, $Q$, take place during a time and length scale $\propto Q^{-1}$ that should not be resolved by the medium. Moreover, recent results \cite{Adam2016} show that there is no strong dependency on the suppression of different hadronic species at high $p_\perp$, thus supporting a picture in which hadrons are formed already outside the medium. As such, even though the medium scale may be below the $\Lambda_{QCD}$, it should be possible to describe observables related to hard processes in heavy-ion collisions within a pQCD approach. Assuming a factorisation scheme in which the medium-modifications are accounted for a modification of the parton branching structure it follows:
\begin{equation}
	d \sigma_{med}^{A A \rightarrow h + rest} = \sum_{ijk} f_{i/A}(x_1, Q^2) \otimes f_{j/A}(x_2, Q^2) \otimes \hat{\sigma}_{ij \rightarrow f+k} \otimes P_f(\Delta E, L, \hat{q}, \cdots) \otimes D_{f \rightarrow h} (z, \mu_F^2) \, ,
\end{equation}
where the modifications to the showering of the parton $f$, $P_f$, will depend on the energy loss, $\Delta E$, and on the several parameters necessary to describe the medium characteristics (medium length, $L$, transport coefficient, $\hat{q}$, ...). 
\par This manuscript is organized as follow: section \ref{sec:vac}, will be dedicated to make a brief overview about the main building blocks that are necessary to build the parton shower evolution in vacuum. This constitutes a well defined baseline to understand the modifications of the parton branching evolution in the presence of a QCD medium. In section \ref{sec:med}, it will be discussed the medium modifications to the elementary processes that build up the in-medium parton shower evolution within a pQCD approach. Main focus will be given to the contribution of the work developed by the author in this manuscript. Final conclusions will follow in section \ref{sec:conclusions}.

\section{Vacuum Parton Shower}
\label{sec:vac}
\begin{figure*}
\centering
  \includegraphics[width=0.6\textwidth]{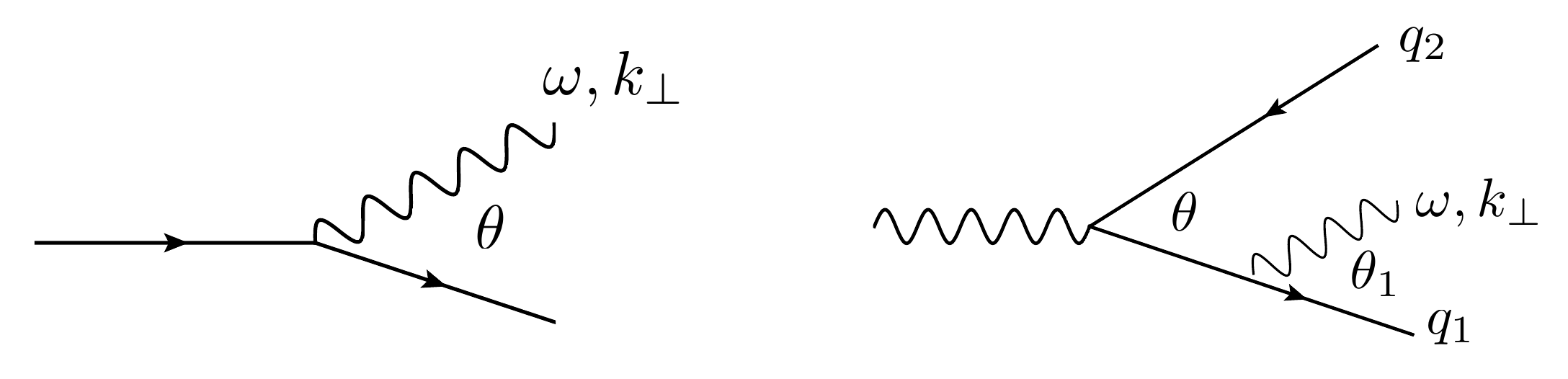}
\caption{Diagrams of the processes that constitute the building blocks of a QCD parton shower: (left) gluon bremsstrahlung; (right) gluon emission from a quark-antiquark antenna.}
\label{fig:buildingBlocks} 
\end{figure*}
\par The pQCD description of a high virtuality parton, in vacuum, is dominated by gluon bremsstrahlung.  The process is illustrated in figure \ref{fig:buildingBlocks} (left). The probability of emitting a gluon of energy $\omega$ and transverse momentum $k_\perp$ is given by:
\begin{equation}
	dP^{q \rightarrow qg} \sim \alpha_s C_R \frac{d\omega}{\omega} \frac{dk_\perp^2}{k_\perp^2} \, ,
\label{eq:bremsstrahlung}
\end{equation}
where $\alpha_s = g^2/(4 \pi)$, $g$ is the coupling constant and $C_R$ the Casimir color factor ($R = F$ for parent quarks and $R = A$ for parent gluons). An evident feature from equation (\ref{eq:bremsstrahlung}) is the double log enhancement for collinear and soft gluon emissions. However, the resummation of such contributions alone fail to describe the jet fragmentation functions by overpopulating the soft modes. To correctly account for multiple gluon emissions, it is necessary to check the interferences between subsequent emissions. It can be shown (see, for instance \cite{Dokshitzer:1991wu,Ellis:1991qj} for a complete review) that such interferences actually reduce the available phase space for subsequent radiation, originating the known property of \textit{angular ordering}. This phenomena constitutes the basis of every Monte Carlo event generator. This can be more easily observed when considering the emission of a soft gluon from a quark-antiquark pair that was initiated by a photon\footnote{When the parent parton of the the quark-antiquark antenna is a gluon, an additional term $\propto C_A \Theta( \cos \theta - \cos \theta_1)$ appears. These emissions, since they are proportional to the color of the gluon, are re-interpreted as emissions from the parent parton before the splitting. As such, the same angular-ordering interpretation holds.} (see figure \ref{fig:buildingBlocks}, right). After integrating over azimuthal angle, the spectrum of radiated gluons of the quark, and similarly for the antiquark, goes as:
\begin{equation}
	dN_{q}^{\omega \rightarrow 0} \sim \alpha_s C_F \frac{d\omega}{\omega} \frac{\sin \theta d\theta}{1 - \cos \theta} \Theta( \cos \theta_1 - \cos \theta) \, ,
\label{eq:vacAntenna}
\end{equation}
where $\theta_1$ is the angle between the radiated gluon and the parent parton and $C_F = (N^2 - 1)/(2N)$ being $N = 3$ the number of colors. The $\Theta$ function implies that $\theta > \theta_1 > \theta_2 > \cdots $, thus suppressing large angle emissions. This effect was experimentally observed by the TASSO \cite{Braunschweig:1990yd} and OPAL \cite{Abbiendi:2002mj} collaborations. 

\section{Medium Parton Shower}
\label{sec:med}

\subsection{In-Medium Propagation}
\label{subsec:formalism}
\par
\begin{figure*}
\centering
  \includegraphics[width=0.35\textwidth]{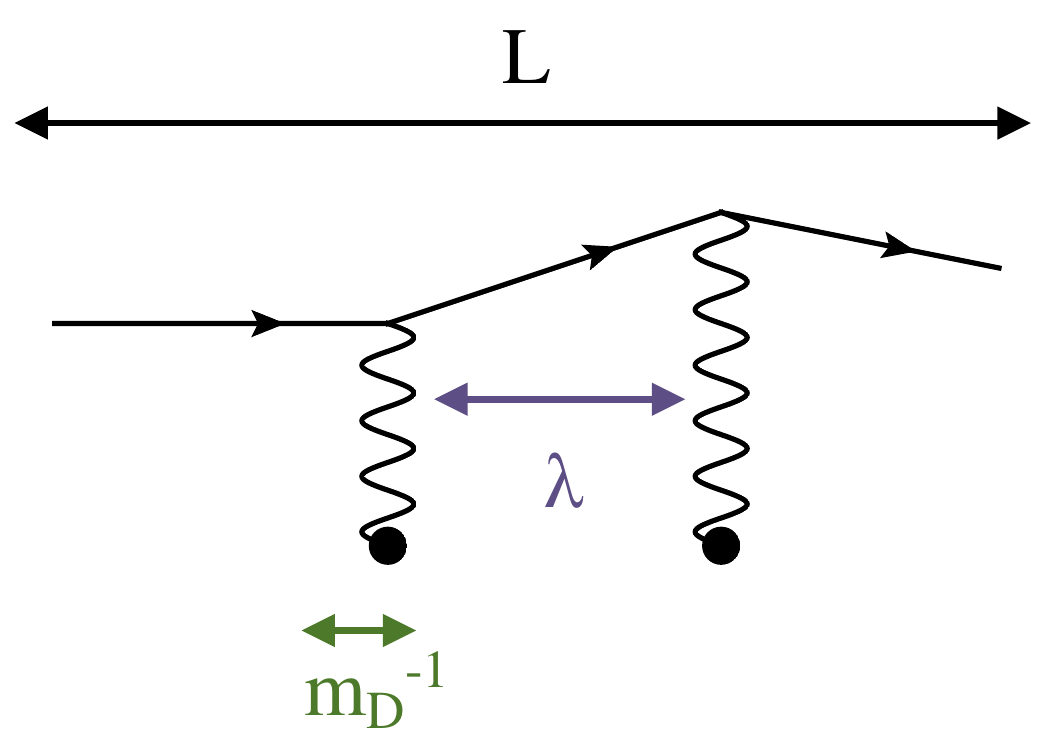}
\caption{Illustration of the ordering of the medium scales ($ L \gg \lambda \gg m_D^{-1}$) to make a perturbative treatment of the medium dynamics and medium-probe interactions. The scattering centres are represented as black dots and the probe as an arrow. Each scattering centre induces a transverse momentum \textit{kick} in the transverse direction.}
\label{fig:medium} 
\end{figure*}

 The considered medium (QGP) is a strongly coupled fluid and, therefore, its interactions are at a scale that is non perturbative. Nonetheless, the evolution of a hard probe (particle or jet) that is propagating though this matter is, as seen in section \ref{sec:vac}, described by pQCD since it has a virtuality scale $\sqrt{Q^2} \gg \Lambda_{QCD}, T$, where $T$ is the temperature of the plasma. In this work, we assume that the medium description and its interaction with a hard probe can be described within such perturbative approximation. As such, the medium is seen as a collection of independent static scattering centres such that $L \gg \lambda \gg m_{D}^{-1}$ where $L$ is the medium length, $\lambda$ the mean-free path of the particle inside the medium and $m_{D}$ the Debye screening mass, see figure \ref{fig:medium}. 
The average squared transverse momentum acquired by the propagating particle, $\left\langle k_\perp^2 \right\rangle$, per each $\lambda$, 
\begin{equation}
	\hat{q} = \frac{\left\langle k_\perp^2 \right\rangle}{\lambda} \, ,
\label{eq:qhat}
\end{equation}
is called the transport coefficient and, together with the medium length $L$, are the only parameters necessary to describe the medium characteristics within the employed approximations.
\par Using the high energy approximation\footnote{In the remaining of this manuscript, it will be used light cone coordinates, $x = (x_+, x_-, x_\perp)$, whose relation with Minkowski coordinates are as follows:  $x_+ = \frac{x_0 + x_3}{\sqrt{2}}$, $x_- = \frac{x_0 - x_3}{\sqrt{2}}$ and $x_\perp = (x_1, x_2)$.}, $k_+ \gg k_\perp$. No energy loss is assumed to occur in the longitudinal direction, and the result of the several interactions with the medium scattering centres is a color phase rotation of the incoming particle, that is described by a Wilson Line:
\begin{equation}
	W_{BA} (x_{0+}, L_+; x_\perp) = \mathcal{P} \exp \left\{ i g \int_{x_{0+}}^{L_+} dx_+ A_- (x_+, x_\perp) \right\} \, ,
\label{eq:WilsonLine}
\end{equation}
where $A$ and $B$ are the initial and final color states of the incoming particle and $A_-$ the medium fields that are path ordered in the longitudinal direction $x_+$ and transverse position $x_\perp$. The longitudinal boundaries of the medium are made explicit in the integral. On top of this modification, the particle receives a transverse momentum \textit{kick} from each interaction. Such effect will induce a Brownian motion in the transverse plane, from $x_{0,\perp}$ at the longitudinal position $x_{0+}$ to $x_\perp$ at $L_+$, that is described by a Green's function:
\begin{equation}
\begin{split}
	G_{BA} (x_{0+}, x_{0,\perp}, L_+, x_\perp | p_+) & = \int_{r_\perp(x_{0+})}^{r_\perp(L_+)} \mathcal{D} r_\perp (\xi) \exp\left\{\frac{i k_+}{2} \int_{x_{0+}}^{L_+} d\xi \left( \frac{dr_\perp}{d\xi} \right)^2 \right\}  \\
	& \times W_{BA}(x_{0+}, L_+; r_\perp(\xi)) \, .
\end{split}
\label{eq:GreenFunction}
\end{equation}
\par Within the considered limit, it is understood that the dynamics that lead to the modifications of the medium colour structure occur in a timescale that is much larger than the propagation time of the penetrating particle. The computation of the resulting modifications can be done assuming a frozen medium color configuration while an average over the medium configurations ensemble should be carried out at the end. Moreover, one can decouple the color from the kinematic information from equation (\ref{eq:GreenFunction}) when calculating the medium averages. The former accounts for the computation of $N$-field correlators (eq. (\ref{eq:WilsonLine})) that are taken up to second order in the fields, $\left\langle A(x_\perp) A(y_\perp) \right\rangle$. The re-exponentiation of the result leads to:
\begin{equation}
	\frac{1}{N} Tr \left\langle W(x_\perp) W^\dagger (y_\perp) \right\rangle = \exp \left\{ -\frac{C_F}{2} \int dx_+ \sigma (x_\perp - y_\perp) n(x_+) \right\} \, , 
\end{equation}
where $n(x_+)$ is the longitudinal density of scattering centres and $\sigma$ the dipole cross-section. The latter, within a multiple soft scattering approximation, is approximated by its short distance term such that:
\begin{equation}
	C_F n(\xi) \sigma (r) \simeq \frac{1}{2} \hat{q} r^2 + \mathcal{O} (r^2 \ln r^2) \, .
\end{equation}
As for the kinematic information, a semi-classical approximation is usually employed to evaluate the corresponding path-integrals. The dominant contribution is given by the classical trajectory that is found from the classical action, $R_{cl} = \int_{x_+}^{y_+} d\xi \mathcal{L}$, with some additional fluctuations:
\begin{equation}
\begin{split}
	G_0 (x_+, x_\perp; y_+, y_\perp) & = \int_{r_\perp (x_+) = x_\perp}^{r_\perp (y_+) = y_\perp} \mathcal{D} r_\perp(\xi) \exp \left\{ \frac{i p_+}{2} \int_{x_+}^{y_+} d\xi \left( \frac{dr_\perp}{d\xi} \right)^2 \right\} \\
	& = \frac{1}{(2 \pi i )^{D/2}} \left| det \left( - \frac{\partial^2 R_{cl}}{\partial y_i \partial x_i} \right)\right|^{1/2} \exp \left\{ i R_{cl} (x_+, x_\perp, y_+, y_\perp) \right\} \, ,
\end{split}
\end{equation}
where $i, j \in \left\{ 1, \cdots, D \right\}$ and $D$ is the number of dimensions.

\subsection{In-Medium Radiation}
\label{subsec:energyLoss}
\begin{figure*}
\centering
  \includegraphics[width=0.5\textwidth]{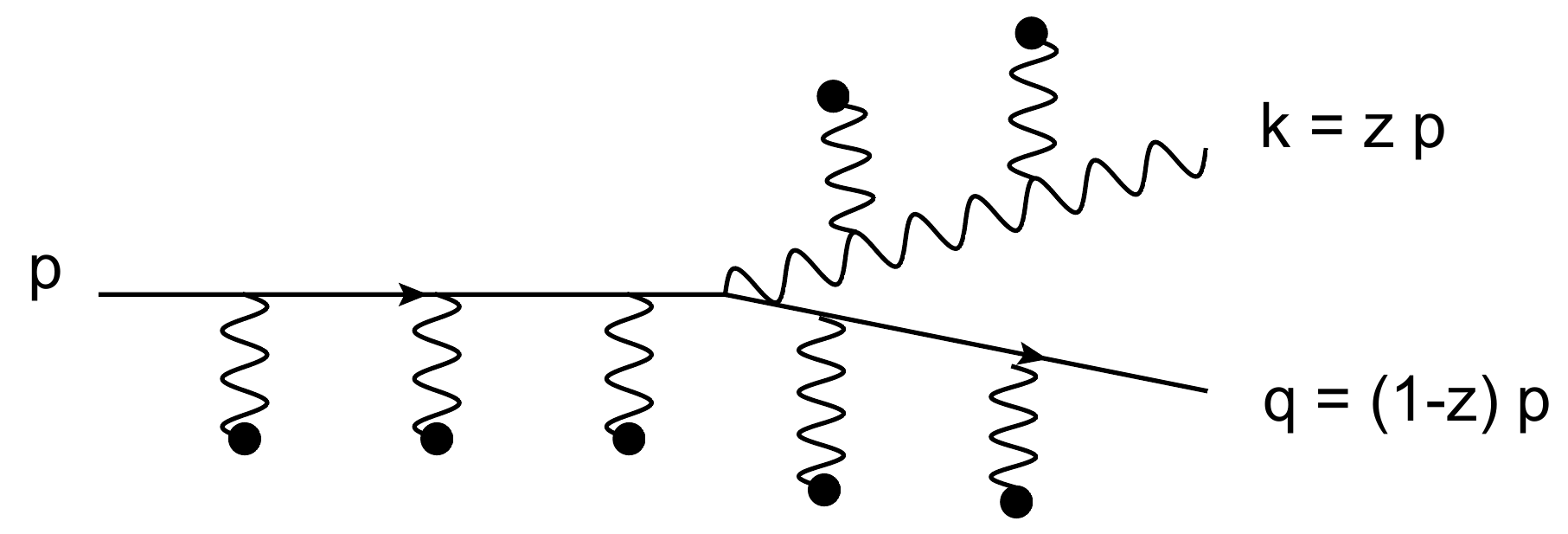}
\caption{Illustration of the in-medium gluon bremsstrahlung process. The initial quark has 4-momenta $p$ and emits a gluon with a fraction $z$ of its longitudinal energy.}
\label{fig:gluonEmission} 
\end{figure*}
\par
\begin{figure*}
\centering
  \includegraphics[width=\textwidth]{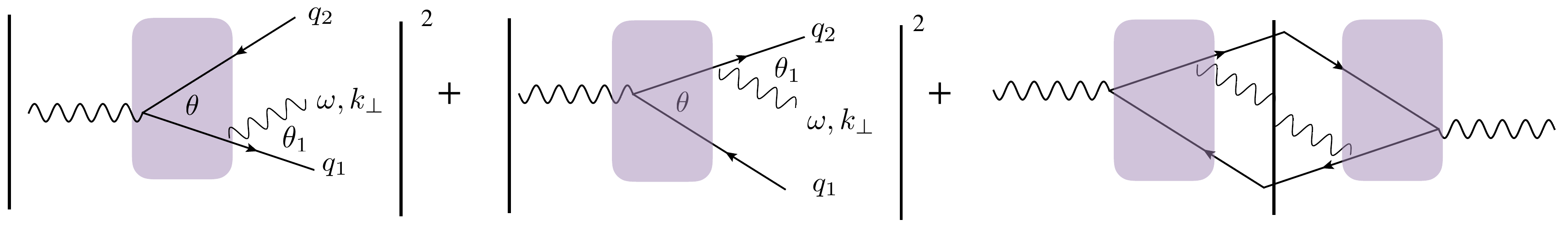}
\caption{Diagrams that contribute to the single-gluon emission energy spectrum from a quark-antiquark antenna setup in a finite medium (represented as the purple region). The 3 contributions correspond to the sum of the three terms in equation (\ref{eq:eikonalAntenna})}
\label{fig:antennaDiagrams}       
\end{figure*}

 Multiple soft scatterings will enhance the resulting gluon radiation due to the accumulation of momenta by each scattering. This process, illustrated in figure \ref{fig:gluonEmission}, has been extensively addressed in the last years (see \cite{Apolinario:2012vy,Arnold:2002ja,Baier1997,Baier1997a,Wiedemann:2000tf,Zakharov1996,Zakharov1997} for works done in the same path-integral approach as this manuscript and \cite{Gyulassy:1999zd,Gyulassy:2000fs} for works done using a Soft and Collinear Effective Theory). More recent works \cite{Apolinario2015,Blaizot2013,Blaizot2014} were able to include all non-eikonal corrections in the description of the in-medium gluon bremsstrahlung. They have shown that the probability of such process is proportional to the independent broadening of the two outgoing particles with a correcting factor \cite{Apolinario2015}:
\begin{equation}
	\Delta_{coh} = 1 + \int_{x_+}^{L_+} d\tau \hat{q} \left. (x_q - x_{\bar{q}}) \cdot (x_{g} - x_{\bar{g}}) \right|_{\tau} \exp \left\{ - \hat{q}_F \int_{x_+}^{\tau} d\xi ( x_{q} - x_{\bar{g}}) \cdot (x_{\bar{q}} - x_g) \right\} \, ,
\label{eq:DeltaCoh}
\end{equation}
where $x_{q(g)}$ stands for the transverse position of the quark (gluon) and in the amplitude and $x_{\bar{q}(\bar{g})}$ in the complex conjugate amplitude. This additional term states that, on top of a complete factorisation of the showering process, there is an additional correction in which both final particles remain in a coherent state. There were several works that address the phenomenology of the parton cascade when neglecting this correction factor \cite{Blaizot2013a,Blaizot2015b,Fister2015,Iancu2015}. In this work, we devoted ourselves to study the consequences of such correction for subsequent emissions. For that, we consider the gluon emission off a quark-antiquark antenna that is propagating through a finite medium. The corresponding diagrams that need to be evaluated are schematically represented in figure \ref{fig:antennaDiagrams}.

\subsection{In-Medium QCD Antenna}
\label{subsec:antenna}
\par The calculation of the diagrams in figure \ref{fig:antennaDiagrams} was done in the eikonal approximation in the works by \cite{Casalderrey-Solana2011,Casalderrey-Solana2013,MehtarTani:2010ma,Mehtar-Tani2012}. The result for the spectrum of radiated gluons can be written as:
\begin{equation}
	\frac{dI}{d\Omega_k} = R_q + R_{\bar{q}} - 2 J (1 - \Delta_{med}) \, ,
\label{eq:eikonalAntenna}
\end{equation}
where the gluon phase space is $d\Omega_k = dk_+/2 k_+ \, d^2k_\perp/(2 \pi)^3$. The Dirac contributions corresponding to the diagrams in figure \ref{fig:antennaDiagrams} are defined as:
\begin{equation}
\begin{split}
	& R_q \sim \alpha_s C_F \frac{q_{1+}}{ (k \cdot q_1) } \ \ \ \ \ \ , \ \ \ \ \ \ R_{\bar{q}} \sim \alpha_s C_F \frac{q_{2+}}{(k \cdot q_2)} \, \\
	& 2J \sim \alpha_s C_F \left[ \frac{q_{1+}}{ (k \cdot q_1) } + \frac{q_{2+}}{(k \cdot q_2)} - \frac{k_+ (q_1 \cdot q_2)}{ (k\cdot q_1) (k\cdot q_2)} \right] \, ,
\end{split}
\label{eq:antennaTerms}
\end{equation}
where:
\begin{equation}
	1 - \Delta_{med} = \frac{1}{N^2} Tr \left\langle W_A (x_q) W^\dagger_A (x_{\bar{q}}) \right\rangle \, ,
\end{equation}
is the non-trivial color structure from the third diagram in figure \ref{fig:antennaDiagrams}. The $A$ subscript refers to the adjoint representation of equation (\ref{eq:WilsonLine}) and it is made explicit the transverse coordinates for the quark ($x_q$) and anti-quark ($x_{\bar{q}}$). Finally, it follows that:
\begin{equation}
	\frac{dI}{d\Omega_k} = R_{coh} + 2 J \Delta_{med} \, ,
\label{eq:eikonalAntenna3}
\end{equation}
where
\begin{equation}
	R_{coh} \sim \alpha_s C_F \frac{k_+ (q_1 \cdot q_2) }{(k\cdot q_1)(k \cdot q_2) } \, ,
\end{equation}
is the result from the \textit{vacuum} antenna (eq (\ref{eq:vacAntenna}) after integrating in azimuthal angle).
\par
\begin{figure*}
\centering
  \includegraphics[width=0.8\textwidth]{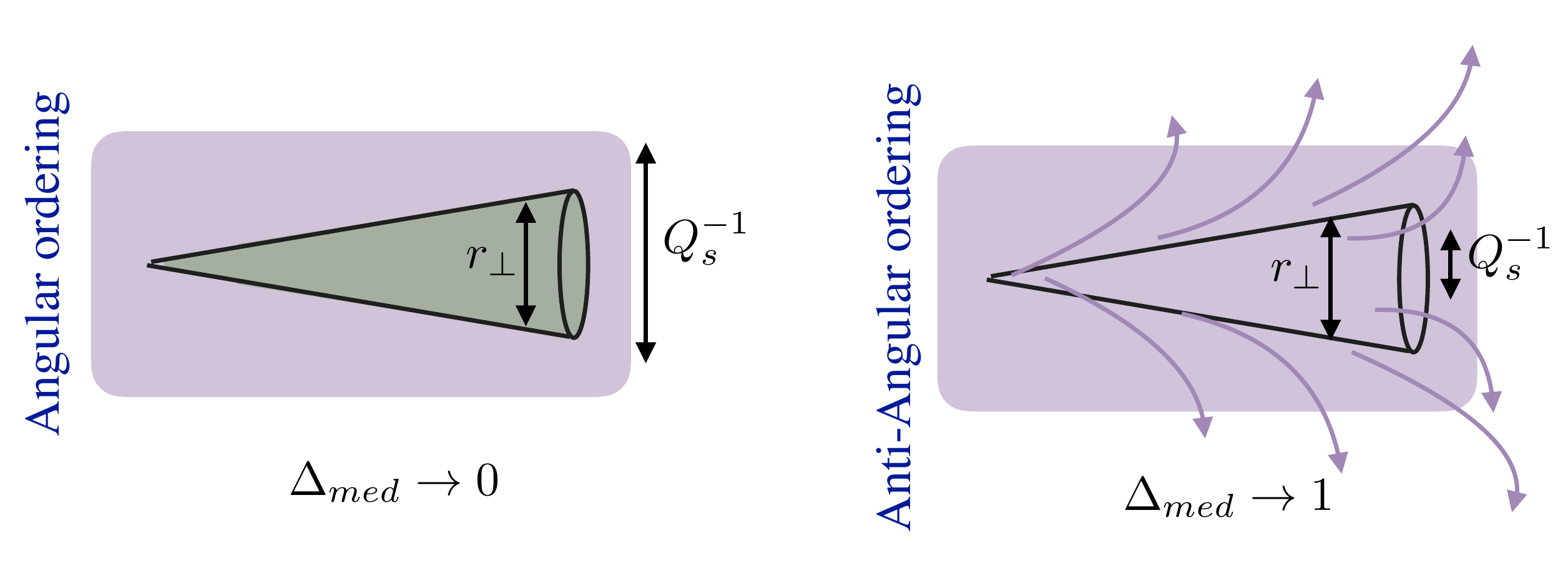}
\caption{Schematic interpretation of the two angular ordering regimes of equation (\ref{eq:eikonalAntenna2}). The left diagram corresponds to $\Theta ( \cos \theta_1 - \cos \theta)$ function, representing the \text{vacuum} angular ordering property. The right diagram corresponds to the $\Theta ( \cos \theta - \cos \theta_1)$ function and stands for the anti-angular ordering property.}
\label{fig:ordering}
\end{figure*}

 After integrating in azimuthal angle, and limiting the present discussion to the soft limit, the spectrum of radiated gluons is proportional to:
\begin{equation}
	dN_q^{\omega \rightarrow 0} \sim \alpha_s C_F \frac{d\omega}{\omega} \frac{\sin\theta d\theta}{1 - \cos \theta} \left[ \Theta( \cos \theta_1 - \cos \theta) + \Delta_{med} \Theta(\cos \theta - \cos \theta_1) \right] \, ,
\label{eq:eikonalAntenna2}
\end{equation}
where (within this limit),
\begin{equation}
	\Delta_{med} \approx 1 - \exp \left\{ - \frac{1}{12} Q_s^2 r_\perp^2 \right\} \, .
\end{equation}
The transverse antenna resolution is $r_\perp = \theta L$ and the medium transverse scale $Q_s^{-1} = (\hat{q} L )^{-1/2}$. From there, it is clear the interplay between these two scales: when $r_\perp \ll Q_s^{-1}$, $\Delta_{med} \rightarrow 0$ and the vacuum result, eq. (\ref{eq:vacAntenna}), where subsequent emissions follow the angular ordering behaviour, is recovered; in the opposite limit, $r_\perp \gg Q_s^{-1}$, $\Delta_{med} \rightarrow 1$ and the available phase space for radiation opens to allow also \textit{anti-angular ordering} emissions, where $\theta < \theta_1 < \theta_2 < \cdots $. A schematic interpretation of this result is shown in figure \ref{fig:ordering} for the two limiting cases: vacuum interferences remain in the development of the shower while the medium is not able to resolve the two emitters independently (figure \ref{fig:ordering}, left); whenever the medium transverse resolution is able to probe quark and anti-quark separately, the destructive interferences are suppressed and a new angular regime becomes available, in which subsequent emissions are emitted at larger angles (figure \ref{fig:ordering}, right).
\par
\begin{figure*}
\centering
  \includegraphics[width=0.8\textwidth]{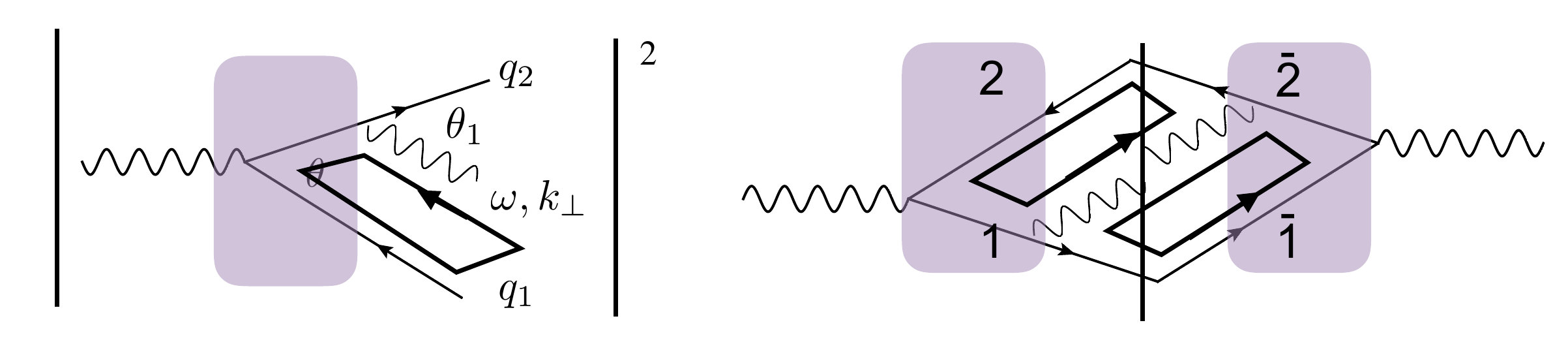}
\caption{Schematic representation of the color structure beyond the eikonal approximation of the diagrams in figure \ref{fig:antennaDiagrams}. The traces in equations (\ref{eq:trace1}) and (\ref{eq:trace2}) are represented as closed arrow loops.}
\label{fig:color}
\end{figure*}

 When going beyond the eikonal limit, the color structure of the diagrams in figure \ref{fig:antennaDiagrams} has to account for transverse Brownian motion, and so, there will be additional terms with respect to the previous case \cite{Mine}. The new non-trivial color structures are represented in figure \ref{fig:color}. As one can see, the direct terms (figure \ref{fig:color}, left) will be proportional to a quadrupole that, within the considered limits (soft gluon radiation and multiple soft scattering approximation), can be approximated by the two independent broadenings of both quark and anti-quark:
\begin{equation}
\begin{split}
	\Delta_{coh}^\prime & = \frac{1}{N2} Tr \left\langle G_1 G_2^\dagger G_{\bar{1}}^\dagger G_{\bar{2}} \right\rangle = \frac{1}{N^2} Tr \left\langle G_1 G^\dagger_2 \right\rangle Tr \left\langle G_{\bar{1}}^\dagger G_{\bar{2}} \right\rangle \Delta_{coh} \\
	& \simeq \frac{1}{N^2} Tr \left\langle G_1 G^\dagger_2 \right\rangle Tr \left\langle G_{\bar{1}}^\dagger G_{\bar{2}} \right\rangle \, ,
\end{split}
\label{eq:trace1}
\end{equation}
where $1 (\bar{1}), 2 (\bar{2})$ refers to the quark and anti-quark in the (complex conjugate) amplitude. As for the interference term (figure \ref{fig:color}, right), a similar $\Delta_{med}$ parameter that includes transverse momentum broadening can be defined such as:
\begin{equation}
	1 - \Delta_{med}^\prime = \frac{1}{N^2} Tr \left\langle G_1 G^\dagger_2 \right\rangle Tr \left\langle G_{\bar{1}}^\dagger G_{\bar{2}} \right\rangle \, .
\label{eq:trace2}
\end{equation}
The Dirac structure will be the same as equation (\ref{eq:antennaTerms}). 
The spectrum in equation (\ref{eq:eikonalAntenna}) (or eq. (\ref{eq:eikonalAntenna3})) is modified such that:
\begin{equation}
	\frac{dI}{d\Omega_q d\Omega_{\bar{q}} d\Omega_k} = \Delta_{coh}^\prime (R_q + R_{\bar{q}}) - 2 (1-\Delta_{med}^\prime) J = \Delta_{coh}^\prime R_{coh} + 2 \overline{\Delta_{med}} J \,
\label{eq:nonEikonalSpec}
\end{equation}
where it was defined $(1- \Delta_{med}^\prime) = \Delta_{coh}^\prime - \overline{\Delta_{med}}$ to have the same diagrammatic interpretation as in figure \ref{fig:ordering}. There is a correction due to the transverse Brownian motion such that the angular-ordering \textit{cone} opens by:
\begin{equation}
	\Delta_{coh}^\prime \sim \frac{4 \pi}{(\hat{q} \Delta L_+)^2 } \left[ (k_\perp + q_{1,\perp} - q_{2,\perp})^2 + \hat{q} \Delta L_+ \right] \exp \left\{ - \frac{ (q_{1,\perp}+q_{2,\perp})^2}{\hat{q} \Delta L_+} \right\} \, ,
\end{equation}
where $\Delta L_+ = L_+ - x_{0+}$.
As for the anti-angular component, instead of straight trajectories, the two emitters will oscillate in the transverse plane with:
\begin{equation}
\begin{split}
	1 - \Delta_{med}^\prime & \sim \exp \left\{ - \frac{i \tan (\Omega \Delta L_+)}{\Omega} \left[ (1 - z) (q_{1,\perp} + k_\perp) - z q_{2,\perp} \right]^2 \right. \\
	& \left. + \frac{i \tan (\Omega^\prime \Delta L_+)}{\Omega^\prime} \left[ (1-z) q_{1,\perp} - z (q_{2,\perp} + k_\perp) \right]^2 \right\} \, ,
\end{split}
\end{equation}
where $\Omega^2 \sim \hat{q} / (z (1-z) p_+)$, being $p_+$ the photon longitudinal energy, $z$ the energy fraction carried by the quark and $\Omega^{\prime 2} = - \Omega^2$. 
\par As such, even when considering non-eikonal corrections, one can understand a medium modified jet as composed by angular ordered mini vacuum-like jets. Each of these hard and collimated structures will behave as a different emitter inside of the jet that will lose energy independently. Therefore, the hard structure of the jet should be unaffected by the medium while the amount of soft fragments should be enhanced with respect to pp collisions. This expectation seems to be in agreement with the latest experimental results from both ATLAS\cite{Aad:2014wha} and CMS\cite{Chatrchyan:2014ava} collaborations.

\section{Conclusions}
\label{sec:conclusions}
\par Significant progress been made to understand the pQCD evolution of the parton shower in the presence of a hot and dense medium. 
In this work, we calculated the medium-induced gluon radiation from a non-eikonal QCD antenna in a finite medium. The results show that it is possible to generalize the previous results derived in a strict eikonal approximation to account for transverse momentum broadening. Nonetheless, the qualitative picture of a medium-modified jet remains to be valid, even when including such non-eikonal corrections.  The resulting picture seems to be also in agreement with experimental observations.
\par Continuously efforts are still necessary to improve the current limitations of the qualitative picture obtained so far. Although it is clear that there is a strong interplay between medium- and vacuum-like jets, it is still unclear how to derive analytical evolution equations for both as the ordering variable by which one is able to resum the dominant contributions is not the same ($t \sim ln Q^2$ for vacuum-like jets while $t \sim L$ for medium-like jets). Moreover, the generalisation to multiple branchings is necessary to accurately evaluate the level of approximation when considering a complete factorised picture during the parton shower evolution. 

\textbf{Acknowledgements}
The author thanks N. Armesto, G. Milhano and C. Salgado for carefully reading this manuscript. This work was funded by the portuguese Funda\c{c}\~{a}o para a Ci\^{e}ncia e Tecnologia, grants SFRH/ BPD/103196/2014 and CERN/FIS-NUC/0049/2015.

\bibliographystyle{ieeetr}
 \bibliography{MyCollection.bib}

\begin{thebibliography}{10}

\bibitem{Adam2016}
{ALICE Collaboration}, ``{Centrality dependence of the nuclear modification
  factor of charged pions, kaons, and protons in Pb-Pb collisions at
  $\sqrt{s_{NN}}$ =2.76 TeV},'' {\em Phys. Rev. C.}, vol.~93, no.~3, p.~034913,
  2016.

\bibitem{Dokshitzer:1991wu}
Y.~L. Dokshitzer, V.~A. Khoze, A.~H. Mueller, and S.~I. Troian, {\em {Basics of
  perturbative QCD}}.
\newblock 1991.

\bibitem{Ellis:1991qj}
R.~K. Ellis, W.~J. Stirling, and B.~R. Webber, {\em {QCD and collider
  physics}}, vol.~8.
\newblock 1996.

\bibitem{Braunschweig:1990yd}
W.~Braunschweig {\em et~al.}, ``{Global Jet Properties at 14-{GeV} to 44-{GeV}
  Center-of-mass Energy in $e^+ e^-$ Annihilation},'' {\em Z. Phys.}, vol.~C47,
  pp.~187--198, 1990.

\bibitem{Abbiendi:2002mj}
G.~Abbiendi {\em et~al.}, ``{Charged particle momentum spectra in e+ e-
  annihilation at s**(1/2) = 192-GeV to 209-GeV},'' {\em Eur. Phys. J.},
  vol.~C27, pp.~467--481, 2003.

\bibitem{Apolinario:2012vy}
L.~Apolinario, N.~Armesto, and C.~A. Salgado, ``{Medium-induced emissions of
  hard gluons},'' {\em Phys. Lett.}, vol.~B718, pp.~160--168, 2012.

\bibitem{Arnold:2002ja}
P.~B. Arnold, G.~D. Moore, and L.~G. Yaffe, ``{Photon and gluon emission in
  relativistic plasmas},'' {\em JHEP}, vol.~06, p.~030, 2002.

\bibitem{Baier1997}
R.~Baier, Y.~Dokshitzer, A.~Mueller, S.~Peign{\'{e}}, and D.~Schiff,
  ``{Radiative energy loss and pT-broadening of high energy partons in
  nuclei},'' {\em Nucl. Phys. B}, vol.~484, pp.~265--282, 1997.

\bibitem{Baier1997a}
R.~Baier, Y.~Dokshitzer, A.~Mueller, S.~Peign{\'{e}}, and D.~Schiff,
  ``{Radiative energy loss of high energy quarks and gluons in a finite-volume
  quark-gluon plasma},'' {\em Nucl. Phys. B}, vol.~483, no.~1, pp.~291--320,
  1997.

\bibitem{Wiedemann:2000tf}
U.~A. Wiedemann, ``{Jet quenching versus jet enhancement: A Quantitative study
  of the BDMPS-Z gluon radiation spectrum},'' {\em Nucl. Phys.}, vol.~A690,
  pp.~731--751, 2001.

\bibitem{Zakharov1996}
B.~G. Zakharov, ``{Fully quantum treatment of the Landau--Pomeranchuk--Migdal
  effect in QED and QCD},'' {\em J. Exp. Theor. Phys. Lett.}, vol.~63, no.~12,
  pp.~952--957, 1996.

\bibitem{Zakharov1997}
B.~G. Zakharov, ``{Radiative energy loss of high-energy quarks in finite-size
  nuclear matter and quark-gluon plasma},'' {\em Jetp Lett.}, vol.~65, no.~8,
  pp.~615--620, 1997.

\bibitem{Gyulassy:1999zd}
M.~Gyulassy, P.~Levai, and I.~Vitev, ``{Jet quenching in thin quark gluon
  plasmas. 1. Formalism},'' {\em Nucl. Phys.}, vol.~B571, pp.~197--233, 2000.

\bibitem{Gyulassy:2000fs}
M.~Gyulassy, P.~Levai, and I.~Vitev, ``{NonAbelian energy loss at finite
  opacity},'' {\em Phys. Rev. Lett.}, vol.~85, pp.~5535--5538, 2000.

\bibitem{Apolinario2015}
L.~Apolin{\'{a}}rio, N.~Armesto, J.~G. Milhano, and C.~A. Salgado,
  ``{Medium-induced gluon radiation and colour decoherence beyond the soft
  approximation},'' {\em J. High Energy Phys.}, vol.~02, no.~2, p.~119, 2015.

\bibitem{Blaizot2013}
J.~P. Blaizot, F.~Dominguez, E.~Iancu, and Y.~Mehtar-Tani, ``{Medium-induced
  gluon branching},'' {\em J. High Energy Phys.}, vol.~01, no.~1, p.~143, 2013.

\bibitem{Blaizot2014}
J.~P. Blaizot, F.~Dominguez, E.~Iancu, and Y.~Mehtar-Tani, ``{Probabilistic
  picture for medium-induced jet evolution},'' {\em J. High Energy Phys.},
  vol.~06, no.~6, p.~075, 2014.

\bibitem{Blaizot2013a}
J.~P. Blaizot, E.~Iancu, and Y.~Mehtar-Tani, ``{Medium-induced QCD cascade:
  Democratic branching and wave turbulence},'' {\em Phys. Rev. Lett.},
  vol.~111, no.~5, pp.~3--6, 2013.

\bibitem{Blaizot2015b}
J.~P. Blaizot, Y.~Mehtar-Tani, and M.~A.~C. Torres, ``{Angular Structure of the
  In-Medium QCD Cascade},'' {\em Phys. Rev. Lett.}, vol.~114, no.~22, pp.~1--4,
  2015.

\bibitem{Fister2015}
L.~Fister and E.~Iancu, ``{Medium-induced jet evolution: wave turbulence and
  energy loss},'' {\em J. High Energy Phys.}, vol.~03, no.~3, p.~082, 2015.

\bibitem{Iancu2015}
E.~Iancu and B.~Wu, ``{Thermalization of mini-jets in a quark-gluon plasma},''
  {\em J. High Energy Phys.}, vol.~10, no.~10, p.~155, 2015.

\bibitem{Casalderrey-Solana2011}
J.~Casalderrey-Solana and E.~Iancu, ``{Interference effects in medium-induced
  gluon radiation},'' {\em J. High Energy Phys.}, vol.~08, no.~8, 2011.

\bibitem{Casalderrey-Solana2013}
J.~Casalderrey-Solana, Y.~Mehtar-Tani, C.~A. Salgado, and K.~Tywoniuk, ``{New
  picture of jet quenching dictated by color coherence},'' {\em Phys. Lett. B},
  vol.~725, no.~4-5, pp.~357--360, 2013.

\bibitem{MehtarTani:2010ma}
Y.~Mehtar-Tani, C.~A. Salgado, and K.~Tywoniuk, ``{Anti-angular ordering of
  gluon radiation in QCD media},'' {\em Phys. Rev. Lett.}, vol.~106, p.~122002,
  2011.

\bibitem{Mehtar-Tani2012}
Y.~Mehtar-Tani, C.~A. Salgado, and K.~Tywoniuk, ``{Jets in QCD media: From
  color coherence to decoherence},'' {\em Phys. Lett. B}, vol.~707, no.~1,
  pp.~156--159, 2012.

\bibitem{Mine}
L.~Apolin{\'{a}}rio, N.~Armesto, J.~G. Milhano, and C.~A. Salgado, ``{In
  Preparation...},'' 2016.

\bibitem{Aad:2014wha}
G.~Aad {\em et~al.}, ``{Measurement of inclusive jet charged-particle
  fragmentation functions in Pb+Pb collisions at $\sqrt{s_{NN}}=2.76 TeV$ with
  the ATLAS detector},'' {\em Phys. Lett.}, vol.~B739, pp.~320--342, 2014.

\bibitem{Chatrchyan:2014ava}
S.~Chatrchyan {\em et~al.}, ``{Measurement of jet fragmentation in PbPb and pp
  collisions at $\sqrt{s_{NN}}=2.76$ TeV},'' {\em Phys. Rev.}, vol.~C90, no.~2,
  p.~024908, 2014.

\end{thebibliography}

%
%
%

\end{document}